\documentclass[10pt]{article}
\usepackage[cp1251]{inputenc}
\usepackage[russian]{babel}
\usepackage{graphicx}

\begin{document}

\twocolumn[\noindent{\small\it  ISSN 1063-7729, Astronomy Reports,
Vol. 51, No. 5, 2007, pp. 327--381. \copyright Pleiades
Publishing, Ltd., 2007. \noindent Original Russian Text \copyright
V.V. Bobylev, A.T. Bajkova, 2007, published in
Astronomicheski$\check{\imath}$ Zhurnal, 2007, Vol. 84, No. 5, pp.
418--428. }

\vskip -4mm

\begin{tabular}{llllllllllllllllllllllllllllllllllllllllllllllll}
 & & & & & & & & & & & & & & & & & & & & & & & & & & & & & & & & & & & & & & & \\
\hline \hline
\end{tabular}

\vskip 2.5cm

\centerline{\LARGE\bf Analysis of the Velocity Field of F and G
Dwarfs }\centerline{\LARGE\bf in the Solar Neighborhood  as a
Function of Age }

\bigskip

\centerline{\bf  V. V. Bobylev and A. T. Bajkova}

\medskip

\centerline{\it Pulkovo Astronomical Observatory, Russian Academy
of Sciences,} \centerline{\it Pulkovskoe sh. 65, St.Petersburg,
196140 Russia} \centerline{\small Received June 18, 2006; in final
form, October 12, 2006}

\vskip 1.5cm

{\bf Abstract} --- {\small The space velocities from the catalog
of Nordstr$\ddot{o}$m et al. (2004) are used to trace variations
of a number of kinematic parameters of single $F$ and $G$ dwarfs
as a function of their age. The vertex deviation of disk stars
increases from $7\pm 1^{\circ}$ to $15\pm 2^{\circ}$ as the mean
age decreases from 4.3 to 1.5 Gyr. The two-dimensional velocity
distributions in the $UV, UW$, and $VW$ planes are analyzed. The
evolution of the main peaks in the velocity distributions can be
followed to an average age of $\approx 9$ Gyr. We find that: (1)
in the distributions of the UV velocity components, stars of
different types are concentrated toward several stable peaks (the
Hyades, Pleiades, and Sirius Cluster), suggesting that the stars
belonging to these formations did not form simultaneously; (2) the
peak associated with the Hyades Cluster dominates in all age
intervals; and (3) the Hyades peak is strongest for stars with an
average age of 1.5 Gyr, suggesting that this peak contains a
considerable fraction of stars from the Hyades cluster. The age
dependences of the kinematic parameters exhibit a break near
$\approx 4.5$ Gyr, which can be explained as an effect of the
different contributions of stars of the thin and thick disks. The
Str$\ddot{o}$mberg relation yields a solar LSR velocity of
$V_{\odot LSR} = (8.7, 6.2, 7.2)\pm (0.5, 2.2, 0.8)$ km/s. }

\bigskip


{\bf DOI}: 10.1134/S1063772907050034

\vskip 1cm

]

\centerline{1.~INTRODUCTION }

\medskip

Analysis of the velocity field for stars in the solar neighborhood
is of great importance for understanding the kinematics and
evolution of various structural components of the Galaxy. As is
now well known, the distribution of stellar space velocities has a
complex small-scale structure, which could be due to various
physical factors. Whereas the statistical method [1], which can
yield such features of the distribution as the dispersion of the
residual velocities and the orientation of the Schwarzschild
ellipsoid, was previously sufficient to describe the observed
velocity field, we must now use more subtle methods (spatially
noninvariant (adaptive) smoothing, as that used by Skuljan et al.
[2], wavelet analysis, etc.) to identify stable structural
formations, e.g., in the form of peaks [2, 3] or branches [2].

The distributions of space velocities of stars in the solar
neighborhood exhibit several characteristic peaks associated with
well-known open clusters [2, 4.6], such as the Pleiades (with an
age of 70. 125 Myr [7]), the Sirius cluster (500 Myr [8]), and the
Hyades (650 Myr [9]). These are fairly young compared to the age
of the Galaxy (of the order of 10 Gyr). Chandrasekhar [10] showed
that the time scale for the stability of an open cluster is an
order of magnitude shorter than the age of the Galaxy. This
imposes constraints on using the theory of streaming motions
[11.13]. It is therefore of great interest to study the kinematic
characteristics of stars as a function of their age.

The high-precision parallaxes and proper motions provided for a
large number of stars by the HIPPARCOS [14] catalog revealed fine
structure in the distribution of space velocities for stars in the
solar neighborhood [2, 3, 15]. However, these were only
preliminary results, since they were either based on modelled
radial velocities [3, 15] or used insufficiently accurate radial
velocities [2]. In this connection, the survey of
Nordstr$\ddot{o}$m et al. [16] is clearly of great value, as it
gives high-precision radial velocities, proper motions, and
parallaxes for a large and homogeneous sample of $F$ and $G$ stars
and, which is very important, reliable estimates of the ages of
individual stars.

The aim of the current paper is to study the structure of the
distribution of space velocities of $F$ and $G$ dwarfs using the
age estimates listed in [16] in order to follow the evolution of
the main peaks associated with known clusters. Looking for
concentrations of stars of different ages toward the same peaks is
also very important because it would imply the action of some
continuously operating gravitational factor  (spiral density
waves, a bar). In particular, Famaey et al. [17] independently
argue for this possibility based on other methods applied to other
stellar samples.

In this paper, we use the statistical method to determine the
elements of the Schwarzschild ellipsoid, and apply an adaptive
Gaussian smoothing method to the initial stellar distributions to
analyze the fine structure of the velocity field.

We also aim to verify the Str$\ddot{o}$mberg relation based on
space velocities obtained through a joint analysis of proper
motions from the HIPPARCOS catalog, high-precision radial
velocities, and stellar age estimates.

\vskip 1cm

\centerline{2.~INITIAL DATA }

\medskip

The catalog [16], which contains about 14 000 $F$ and $G$ dwarfs,
gives original high-recision radial velocities (with typical
errors of $\approx 0.25$ km/s), previously published $uvby\beta$
photometry in the Str$\ddot{o}$mgren system, HIPPARCOS parallaxes,
supplemented in a number of cases by photometric distances, and
stellar proper motions from the HIPPARCOS and TYCHO- 2 [18]
catalogs. Most of the catalog stars have age estimates determined
with typical accuracies of < 50\%. We considered only single stars
located within 200 pc of the Sun. We did not consider double and
multiple stars marked by a flag in column 4 of the table in
[16].We used only stars with age estimates $\tau$. We corrected
the radial velocities and proper motions for Galactic rotation
using the Oort constants $A = 13.7$ km/s$\cdot$kpc and $B = -12.9$
km/s$\cdot$kpc [19]. In the first part of this paper, we analyze
the nearest single stars for which

$$
 \displaylines
{\hfill
 e_\pi/\pi<0.1,~~~~e_\tau/\tau<0.3. \hfill \llap(1)
}
$$
These are stars with the best age and distance estimates. We need
a large number of such stars to construct two-dimensional
velocity-field distributions of suitable quality. To this end, we
used slightly more distant single stars satisfying the conditions

$$
  \displaylines
{\hfill
 e_\pi/\pi<0.2,\hfill\llap(2)\cr\hfill
 e_\tau/\tau<0.5,\hfill\cr\hfill
 |V_{pec}|<100~{\hbox { km/s}}, \hfill
}
$$
where $V_{pec}$ is the peculiar velocity of a star relative to the
Local Standard of Rest (LSR). We adopted the solar velocity
relative to the LSR presented by Dehnen and Binney [15]:
$(X_\odot,Y_\odot,Z_\odot)_{\hbox {\rm\tiny МСП}}=
 (10.00\pm 0.36,5.25\pm 0.62,7.17\pm 0.38)$~km/s.

\vskip 1cm

\centerline{3.~ANALYSIS METHODS}

\medskip

\centerline{\it 3.1. Statistical Method}

\medskip

We use here a Cartesian coordinate system with axes directed away
from the observer and toward the Galactic center ($l = 0^\circ, b
= 0^\circ$, the $x$ axis), in the direction of Galactic rotation
($l = 90^\circ, b = 0^\circ$, the $y$ axis), and toward the North
Galactic Pole ($b = 90^\circ$, the $z$ axis).

We determined the elements of the Schwarzschild ellipsoid using
the well-known statistical method described in detail in [1,
20--22]. We first determined the components of the Solar velocity
by solving the Kovalsky.Arie equations, then determined the
elements of the residual-velocity dispersion tensor from a
least-squares solution to the system of equations in six unknowns.
Analysis of the eigenvalues of the velocity-dispersion tensor
yields the principal axes of the residual-velocity ellipsoid,
which we denote $\sigma_{1,2,3}$, and the directions of the
principal axes, which we denote $l_{1,2,3}, b_{1,2,3}$. We use the
method of Parenago [23] with fourth-order moments to estimate the
errors in the parameters $\sigma_{1,2,3}$.

\medskip

\centerline{\it 3.2. Constructing the Two-Dimensional Stellar}
\centerline{\it Velocity Distributions }

\medskip

We estimated the two-dimensional probability density $f(U, V)$
based on the computed, discrete $UV$ velocity components using an
adaptive-smoothing method. Unlike Skuljan et al. [2], we used the
radially-symmetric Gaussian kernel

$$
K(r)=\frac{1}{\sqrt{2\pi}\sigma} \exp{-{\frac{r^2}{\sigma^2}}},
$$

where $r^2=x^2+y^2$. This function obeys the condition $\int
K(r)dr=1$, which is necessary for estimating the probability
density. The typical uncertainties in the velocities in our case
is 2 km/s, which, among other factors, determined our choice of
discretization interval for the two-dimensional maps; the area of
each square pixel is $s=2\times 2$km$^2$/s$^2$.

The main idea of the method consists in the following. At each
point, we smooth with a beam whose size is determined by $\sigma$
and varies in accordance with the data density in the neighborhood
of the point considered. Thus, the smoothing is performed with a
relatively narrow beam in regions with high data densities, with
the beam width increasing as the data density decreases.

We used the following form of adaptive smoothing at an arbitrary
point $\xi=(U,V)$ [2]:

$$
 \hat{f}(\xi)=\frac{1}{n}\sum_{i=1}^n K\left(\frac{\xi-\xi_i}{h\lambda_i}\right),
$$
where $\xi_i=(U_i,V_i)$, $\lambda_i$ is the local dimensionless
beam parameter at $\xi_i$, $h$ is a general smoothing parameter,
and $n$ is the number of data points $\xi_i=(U_i,V_i)$. The
parameter $\lambda_i$ of the two-dimensional plane UV is
determined at each point as follows:

$$
 \lambda_i=\sqrt\frac{g}{\hat{f}(\xi)}~,~~~~~
 \ln g=\frac{1}{n}\sum_{i=1}^n \ln \hat{f}(\xi),
$$
where $g$ is the geometric mean of $\hat{f}(\xi)$.

It is obvious that, to determine $\lambda_i$, we must know the
distribution $\hat{f}(\xi)$, which, in turn, can be determined if
all the $\lambda_i$ are known. Therefore, the sought for
distribution must be reconstructed iteratively. As a first
approximation, we used the distribution obtained by smoothing the
initial $UV$ map using a fixed-sized beam. We found the best-fit
value of $h$ by minimizing the mean squared residual for the
difference between the estimated distribution $\hat{f}(\xi)$ and
true distribution $f(\xi)$, which was equal to about 9 km/s in our
case.

\vskip 1cm

\centerline{4.~RESULTS AND DISCUSSION}

\medskip

\centerline{\it 4.1.~Asymmetric Drift}

\medskip

Figures 1 and 2 show the group velocities of stars (with reversed
sign) as functions of the square of the residual-velocity
dispersion $S^2$ and of the stellar age $\tau$. Like Dehnen and
Binney [15], we derived the dependences in Fig.1 using the
quantity $S^2=1.14\cdot\sigma^2_{xx}$. The horizontal bars in Fig.
1 show the one-sigma uncertainties in $S^2$, and the similar bars
in Fig. 2 show the boundaries of the age intervals corresponding
to the mean error of the stellar age determination.

The dependence
$$
 \displaylines
{\hfill
 Y_\odot=a\cdot S^2+b, \hfill \llap(3)
}
$$
shown in the middle plot in Fig. 1 (the Str$\ddot{o}$mberg
relation) has the parameters  $a=0.0122\pm0.0019$~(km/s)$^{-1}$
and $b=6.2\pm2.2$~km/s. We obtained these values using all the
available data.

Extrapolating to zero velocity dispersion yields for the LSR
velocity of the Sun  $Y_{\odot~{\hbox {\rm\tiny МСП}}}=6.2\pm2.2$
km/s. The relatively large error of this velocity is due (middle
plot in Fig. 1) to the substantial scatter in the data points near
$S^2=1000$. As is also evident from the middle plot in Fig. 2, out
to an age of $\tau=4-5$ Gyr, virtually all the points $Y_{\odot}$
are almost parallel to the horizontal axis, and the dependence
shows a peculiar kink at $\tau=4-5$ Gyr. Figure 4 from Dehnen and
Binney [15], which exhibits a depressed portion of the $V$
dependence for velocity dispersions $S^2\approx300-700$(km/s)2,
shows a similar pattern, as does Fig. 6 from Olling and Dehnen
[24] in the color index interval $(B-V)_\circ\approx0-0.3$. After
dropping two data points $Y_{\odot}$in Fig. 1 that correspond to
ages of $\tau=3.7$  and 5.1 Gyr, i.e., to the depressed portion of
the plot, we obtained for the unknown parameters
$a=0.0125\pm0.0009$ (km/s)$^{-1}$ and $Y_{\odot~{\hbox {\rm\tiny
МСП}}}=7.1\pm1.1$ km/s; the accuracy is now twice as good as in
the previous case.

We thus conclude that the effect of this feature is significant
and not accidental. It may be that the form of model relation (3)
used to describe the actual kinematics of stars must be more
complex, since this feature is difficult to explain as an effect
of random errors in the stellar ages. Alternatively, we may
observe a discontinuity due to stars belonging to the thick or
thin disk. We discuss this possibility below in Section 3.3.
Overall, our $a$ and $Y_{\odot}$ values agree with the results of
Dehnen and Binney [15], who obtained the estimates $a=0.0125$
(km/s)$^{-1}$ and $Y_{\odot~{\hbox {\rm\tiny LSR}}}=5.25\pm0.62$
km/s, based on the proper motions of a much greater number of
stars.

We calculated the other two components of the solar velocity by
averaging the velocities of seven stars, without the youngest
stars: $X_{\odot {\hbox {\rm\tiny LSR}}}=8.7\pm0.5$ km/s and
$Z_{\odot {\hbox {\rm\tiny LSR}}}=7.2\pm0.8$ km/s. The
corresponding means in Fig. 1 are indicated by dashed lines.

Equation (3) can be written $Y_\odot=V_\varphi+Y_{\odot~{\hbox
{\rm\tiny LSR}}}$, where $V_\varphi$ is the average velocity lag
of stars with respect to the circular velocity of Galactic
rotation at the distance of the Sun from the Galactic center
($R_o$= 8 kpc), i.e., the asymmetric drift velocity. To get some
idea of the expected dependence $Y_\odot$ on time, we can
substitute the parameters derived from $\sigma=c\cdot \tau^\gamma$
into (3), where $\sigma\equiv S$. We find that the expected
dependence has the form of a power-law function with index
$\gamma=0.66$. We derived the dependence shown by the dashed line
in the middle plot in Fig. 2 directly from the initial data; its
index is $\gamma=0.41\pm0.10$. If we drop the two data points
corresponding to ages $\tau=3.7$  and 5.1 Gyr, we obtain the index
$\gamma=0.37\pm0.06$.

The theory of the dynamic evolution of the Galaxy [25] assumes
that the dependence $V_\varphi$ obeys the relation
$$
 \displaylines
{\hfill
 V_\varphi={\sigma^2_U\over 2 V_{\circ~{\hbox {\tiny LSR}}}}
 \Biggl\{
            {R\over \rho}{d\rho\over dR} +
     2\times{R\over \sigma_U}{d\sigma_U\over dR}\hfill \llap(4) \cr
            + \biggl(1-{\sigma^2_V\over \sigma^2_U} \biggr)+
            \biggl(1-{\sigma^2_W\over \sigma^2_U} \biggr)
 \Biggr\},
}
$$
where ${d\rho/dR}$  is the stellar-density gradient, $R$ is
Galactocentric distance, and $V_{\circ~{\hbox {\rm\tiny LSR}}}$ is
the rotation velocity of the local standard of rest.

The solid line and circles in Fig. 2 show the asymmetric drift
velocity for disk stars. We drew this curve based on the results
of Robin et al [26], which they obtained using formula (4)
together with modern data on the distribution of stellar masses in
the solar neighborhood and estimates of the velocity dispersions
of HIPPARCOS stars as a function of age [27]. We also used the
parameter values [26]  $V_{\circ~{\hbox {\rm\tiny LSR}}}=226$
km/s, $Y_{\odot~{\hbox {\rm\tiny LSR}}}=6.3$ km/s (we merged the
age intervals 5.7 and 7.10 Gyr from Table 4 of  Robin et al. [26]
into a single interval). As is clear from Fig. 2, the dotted and
solid lines, obtained using the different methods, are in good
agreement.

\medskip

\centerline{\it 4.2. Analysis of Two-Dimensional Velocity}
\centerline{\it Distributions as a Function of Age}

\medskip

To construct the velocity distributions in the UV , UW, and VW
planes, we used single stars meeting criteria (2). Unlike the
previous case, the number of stars is 4880, and we subdivided them
into four age groups containing approximately equal number of
stars, which we refer to as t1--t4.

Figures 3.5 show the maps of the $UV, UW$, and $VW$ velocity
distributions for the age samples indicated above.We constructed
all these maps using the adaptive smoothing algorithm (see Section
2.2). The lowest contour level and contour increment are 10\% of
the peak value in all figures. Tables 1 and 2 list the main
kinematic parameters of the samples considered. Table 3 lists the
relative values of the main peaks found in the maps. The
coordinates of these peaks relative to the LSR are equal to (in
km/s)$U= 20$, $V=  8$ for the Sirius peak; $U=  0$, $V= -2$ for
the Coma peak; $U= -5$, $V=-16$ for the Pleiades peak; and
$U=-26$, $V=-12$ for the Hyades peak.

As is clear from Table 1, the behavior of the variation of
$Y_\odot$ for the samples t1--t4 exactly matches that shown in
Fig. 1: the $Y_\odot$ values barely increase out to an age of
$\tau=4.3$ Gyr, and are equal to about 15 km/s. The speed
$V_\odot$ also remains constant and equal to about 19 km/s.

As is evident from Table 2, the third axis $b_3$ does not deviate
significantly from the direction toward the Galactic pole; the
vertex deviation $l_1$ increases from $7\pm1^\circ$ for old stars
to $15\pm2^\circ$ for stars with a mean age of 1.5 Gyr. This
result agrees well with the conclusions of Dehnen and Binney [15],
who found the vertex deviation for main-sequence HIPPARCOS stars
as a function of color index.

The random errors in the stellar radial velocities decrease from
$\approx2$ km/s for the youngest stars to $\approx0.25$ km/s for
old stars. Therefore, the most reliable maps are those for the
samples t2--t3, since, other conditions being the same, they have
lower random space-velocity errors. Our Monte-Carlo numerical
simulations showed that random errors in the stellar space
velocities of 2 km/s shift the maxima of the $UV$ velocity
distributions by no more than $3\div4$ km/s, indicating the
stability of the derived coordinates of the peaks.

It is clear from Figs. 3--5 and Table 3 that concentrations of
stars in the form of stable peaks show up mostly in the
distribution of the UV velocity components (Fig. 3). The Hyades
peak dominates in all age intervals. The relative intensity of the
Hyades peak ($U=-26$ km/s) for young stars is so high that we had
to exclude the youngest stars when determining the solar-velocity
vector relative to the LSR (see the dependences for $X_\odot$ in
Figs. 1.2). This leads us to conclude that the kinematics of the
youngest stars is determined by their membership in the Hyades
peak.

We found no significant variations of the coordinates of the main
peaks as a function of stellar age in samples t1--t4. Table 3
shows the variations of the relative intensities of the peaks.

It is evident from Fig. 3 that the orientations (vertex
deviations) of individual isolated peaks, e.g., of the Hyades
peak, vary appreciably as a function of the sample age. At the
same time, the Hyades and Pleiades peaks form a branch-shaped
extended structure whose orientation remains unchanged. See
Skuljan et al. [2] for a detailed description of such structures
in the $UV$-velocity distribution for a large number of HIPPARCOS
stars.

Figure 3 also demonstrates the development of a structural feature
centered on $U=-25$ and $V=-40$ km/s, which is usually associated
with the $\zeta$ stream. The core of this feature shows up in all
the samples, but it is branch-shaped only for sample t4, i.e., in
the velocity distribution for the oldest stars.

Numerical simulations of disk heating by spiral waves carried out
De Simone et al. [28] showed that the division of the $UV$
distribution into branches and peaks can be explained by
irregularities in the Galactic potential, but not by
irregularities in the star formation processes. The effect of the
bar on the evolution of the velocity ellipsoid and the
distribution of residual stellar velocities are now topics of
extensive studies [29, 30]. Babusiaux and Gilmore [31] provide
strong arguments supporting the presence of a bar, based on an
analysis of infrared observations of stars. Fux [29] showed that a
bar in the Galactic center should result in the development of
arms, and the Hercules arm is believed to be due to the effect of
the bar [17, 32, 33].

Analysis of the $UW$ distribution for sample t1 in Fig. 4 shows
that the Hyades peak shows up conspicuously, along with the
central peak. In the case of sample t4, which contains the oldest
stars, the Hyades peak shows up as a prominent, isolated clump.

The $VW$ velocity distributions (Fig. 5) for samples t1.t4 are
fairly symmetric and regular. The oldest stars (t4) show a
well-defined velocity lag and the ellipse-truncation effect.

On the whole, we conclude that stars of very different ages are
concentrated toward several peaks associated with known open
clusters. Our results agree with the conclusions of Famaey et al.
[17], who selected samples of $M$ and $K$ giants belonging to
individual peaks in the $UV$ velocity plane and computed the
isochrone ages of individual stars. They found that peaks
contained stars covering very wide age intervals, suggesting that
stars belonging to individual peaks did not form simultaneously.
This is also the main conclusion of our paper.

\medskip

\centerline{\it 4.3. Stars of the Thin and Thick Disks}

\medskip

The samples t1--t4 contain stars of both the thin and thick disk,
and may also be slightly contaminated by halo stars. In this
section, we subdivide stars in samples t1--t4 into thin-disk and
thick-disk stars based on kinematic criteria, and analyze the age
dependences of the group velocities of these stars. According to
modern concepts, the conditions $|V_{pec}|<$100 km/s and
[Fe/H]>$-$0.5 dex separate out halo stars rather efficiently [34,
35].We selected a group of single stars meeting the criteria
$$
  \displaylines
{\hfill
 e_\pi/\pi<0.2,~~e_\tau/\tau<0.5,\hfill 
 \cr\hfill
 |V_{pec}|<60~{\hbox {\rm км/с}},~~[Fe/H]>-0.5 {\hbox {\rm~dex}},\hfill
}
$$
which we consider to be thin-disk stars. The second group, which
contains mostly thick-disk stars, satisfies the conditions
$$
  \displaylines
{\hfill
 e_\pi/\pi<0.2,~~e_\tau/\tau<0.5,\hfill 
 \cr\hfill
 60~{\hbox {\rm км/с}}<|V_{pec}|<100~{\hbox {\rm км/с}},\hfill\cr
 \hfill
 [Fe/H]>-0.5 {\hbox {\rm~dex}}.\hfill
}
$$
The calculated kinematic parameters are listed in Table 1, where
the thin-disk and thick-disk samples are indicated by a single and
double prime, respectively; we calculated $V_{pec}$ relative to
the Sun. We show the  age dependences of the resulting
solar-velocity components in Fig. 6. Analysis of Table 1 and Fig.
6 shows that the $Z_\odot$ component shows the smallest
differences between the two groups. The $X_\odot$ velocity
component shows the greatest difference, which reaches 35 km/s for
the youngest stars.

We found the velocity lags at an age of about 9 Gyr to be
$V_{\phi}$=11 km/s for thin-disk stars, reaching $V_{\phi}$=35
km/s for thick-disk stars. These values are consistent with those
currently taken to be the known kinematic characteristics of the
thin and thick disks [24, 26].

It is obviously impossible to separate the evolution of stars of
the thick and thin disks. However, as is clear from Table 1, the
number of thick-disk stars increases appreciably with the sample
age, resulting in the observed discontinuities in the dependences
of  $Y_\odot$ on $\tau$ and on $S^2$. This bend in the dependences
of the kinematic parameters near $\approx4$ Gyr (Fig. 1, 2) can be
explained by changes in the contributions of the thin-disk and
thick-disk stars with stellar age.

It is evident from the middle part of Table 1 and the upper plot
in Fig. 6 that $X_\odot$ for disk stars does not remain constant,
i.e., there is a significant velocity gradient as a function of
time, with this velocity reaching $3.0\pm0.8$ km/s for sample
t4$'$ (or, in terms of signed quantities, ${\overline
U}=-3.0\pm0.8$ km/s).We particularly point out this result,
because, so far, only two large catalogs of high-precision stellar
radial velocities measured with CORAVEL spectrometers are
available: those of Nordstr$\ddot{o}$m et al. [16] and Famey et
al. [17] for dwarf and giant stars, respectively. In their
analysis of high-precision space velocities of $K$ and $M$ giants
in a sample of stars without high-velocity objects, Famey et al.
[17] found ${\overline U}=-2.78\pm1.07$ km/s. This shows that an
appreciable fraction of stars in the solar neighborhood may have
systematic motions in the radial direction (along the Galactic
radius vector), and also further complicates the problem of
selecting stellar samples for the most rigorous determinations of
the parameters of the local solar motion relative to the LSR.

\vskip 1cm

\centerline{CONCLUSIONS }

\medskip

We have used high-precision space velocities (with an average
error of 2 km/s) for single F and G dwarfs within 200 pc of the
Sun taken from the survey of Nordstr$\ddot{o}$m et al. [16] to
analyze the variation of the kinematic parameters of stars as a
function of their age.

We find that the vertex deviation for disk stars increases from
$7\pm1^\circ$ to $15\pm2^\circ$ as the mean age decreases from 4.3
to 1.5 Gyr.

We analyzed the main peaks in the two-dimensional stellar space
velocity distributions in the $UV, UW$, and $VW$ planes associated
with known clusters, to determine how these peaks evolve with
increasing age of the stellar sample, up to a limiting mean age of
$\approx9$ Gyr. This analysis shows the following.

(1) In the $UV$-velocity distribution, stars with different ages
are concentrated toward several stable peaks (the Hyades,
Pleiades, and Sirius clusters). This indicates that stars
belonging to these individual peaks did not all form
simultaneously. This is the main conclusion of this work.

(2) The peak associated with Hyades cluster is the most
conspicuous in all the age intervals.

(3) The Hyades peak is most prominent for stars with a mean age of
1.5 Gyr, suggesting that this peak contains a large fraction of
Hyades cluster stars.

We show that the bend in the age dependences of the kinematic
parameters near $\approx4-5$ Gyr can be explained as an effect of
the changing contributions of thin-disk and thick-disk stars. When
redetermining the parameters of the asymmetric drift and the
Str$\ddot{o}$mberg relation, we found the dependence of $Y_\odot$
on the mean stellar age $\tau$ to show a discontinuity at
$\tau\approx 5.1$ Gyr, with $Y_\odot$ remaining approximately
constant at lower ages ($\approx15$ km/s). Removing outliers
enabled us to increase the accuracy of the parameters of the
Str$\ddot{o}$mberg relation by almost a factor of two, suggesting
that the discontinuity is real and not accidental. Unlike the
well-known Parenago discontinuity, which results from the
subdivision of stars into objects of the disk, intermediate
component, and halo, this discontinuity is due to a subtler
effect: stars of the thick and thin disks contribute differently
to the determination of the kinematic parameters. At $\tau\approx$
9  Gyr, the velocity lags (asymmetric drifts) of the thin and
thick disks are $V_{\phi}$=11 and $V_{\phi}$=35 km/s,
respectively.

The mean velocity component along the $x$ coordinate averaged over
all the stars remains constant, and equal to $X_{\odot~{\hbox
{\rm\tiny LSR}}}=8.7\pm0.5$ km/s. Imposing constraints on
$|V_{pec}|$ leads to the appearance of appreciable gradient of
this quantity as a function of time, and this velocity reaches
$3.0\pm0.8$ km/s for the oldest disk stars. The $z$ velocity
component for all the stars considered is the most stable, and is
equal, on average, to $Z_{\odot~{\hbox {\rm\tiny LSR}}}=7.2\pm0.8$
km/s. Our extrapolation of the residual-velocity dispersion to
zero yielded $Y_{\odot~{\hbox {\rm\tiny LSR}}}=6.2\pm2.2$ km/s.

\vskip 1cm

\centerline{ACKNOWLEDGMENTS}

\medskip

We thank the reviewer for valuable comments that improved the
paper, and to R.B. Shatsova and G.B. Anisimova for calling our
attention to the problem of the discontinuity of the
Str$\ddot{o}$mberg asymmetry. This work was supported by the
Russian Foundation for Basic Research (project code 05-02-17047).

\vskip 1cm

\centerline{REFERENCES}

\medskip
1.~K.~Schwarzschild, Nachr. K\"oniglichen Ges. Wiss. G\"ottingen,
{\bf 191} (1908).

2.~J.~Skuljan, J.~B.~Hearnshaw, and P.~L.~Cottrell, MNRAS {\bf
308}, 731 (1999).

3.~W.~Dehnen, Astron. J. {\bf 115}, 2384 (1998).

4.~E.~Chereul,  M.~Cr\'ez\'e,  and O.~Bienaym\'e,
 Astron. Astrophys. {\bf 340}, 384 (1998).

5.~E.~Chereul , M.~Cr\'ez\'e,  and O.~Bienaym\'e {\it et al.},
Astron. Astrophys. Suppl. Ser. {\bf 135}, 5 (1999).

6.~R.~Asiain, F.~Figueras, J.~Torra {\it et al.}, Astron.
Astrophys. {\bf 341}, 427 (1999).

7.~D.~R.~Soderblom, B.~F.~Jones, S.~Balachandran {\it et al.},
Astron. J.  {\bf 106}, 1059 (1993).

8.~J.~R.~King, A.~R.~Villarreal, D.~R.~Soderblom {\it et al.},
Astron. J. {\bf 125}, 1980 (2003).

9.~V.~Castellani, S.~Degl'Innocenti, and P.~Moroni, MNRAS {\bf
320}, 66 (2001).

10.~S.~Chandrasekhar, {\it Principles of stellar dynamics} (Yerkes
Observatory, 1942; IL, Moscow, 1948).

11.~J.~C.~Kapteyn, British Assoc. Adv. Sci. Rep., 257 (1905).

12.~O.~J.~Eggen,  Astron. J.  {\bf 111}, 1615 (1995).

13.~O.~J.~Eggen,  Astron. J.  {\bf 112}, 1595 (1996).

14.~ESA SP-1200, The Hipparcos and Tycho Catalogues (1997).

15.~W.~Dehnen and J.~J.~Binney, MNRAS {\bf 298}, 387 (1998).

16.~B.~Nordstr\"om, M.~Mayor, J.~Andersen
 {\it et al.}, Astron. Astrophys. {\bf 419}, 989 (2004).

17.~B.~Famaey, A.~Jorissen, X.~Luri {\it et al.},  Astron.
Astrophys. {\bf 430}, 165 (2005).

18.~E.~H{\o}g, C.~Fabricius, V.V.~Makarov  {\it et al.}, Astron.
Astrophys. {\bf 355}, L27 (2000).

19. V. V. Bobylev, Pis'ma Astron. Zh. {\bf 30}, 185 (2004)
[Astron. Lett. {\bf 30}, 159 (2004)].

20. K. F. Ogorodnikov, Dynamics of Stellar Systems (Fizmatgiz,
Moscow, 1958; Pergamon, Oxford, 1965).

21. P. P. Parenago, Kurs zvezdnoi astronomii (Course of Stellar
Astronomy) (Gosizdat, Moscow, 1954) [in Russian].

22. R. J. Trumpler and H. F.Weaver, Statistical Astronomy (Univ.
of Calif. Press, Berkely, 1953).

23. P. P. Parenago, Tr. Gos. Astron. Inst. im. P.K. Sternberga
{\bf 20}, 26 (1951).

24.~R.~Olling and W.~Dehnen, Astroph. J. {\bf 599}, 275 (2003).

25.~I.~R.~King, {\it An Introduction to Classical Stellar
Dynamics} (University of California, Berkeley, 1994; USSR, Moscow,
2002).

26.~A.~C.~Robin, C.~Reyl\'e, S.~Derri\`ere {\it et al.}, Astron.
Astrophys. {\bf 409}, 523 (2003). 

27.~A.~E.~G\'omez, S.~Grenier, S.~Udry {\it et al.}, ESA SP-402:
Hipparcos-Venice' 97, 402 (1997).

28.~R.~S.~De Simone, X.~Wu, and S.~Tremaine, MNRAS {\bf 350}, 627
(2004).

29.~R.~Fux, Astron. Astrophys. {\bf 373}, 511 (2001).

30.~G.~M\"uhlbauer and W.~Dehnen, Astron. Astrophys. {\bf 401},
975 (2003).

31.~C.~Babusiaux and G.~Gilmore, MNRAS {\bf 358}, 1309 (2005).

32.~W.~Dehnen, Astroph. J. {\bf 524}, L35 (1999).

33.~W.~Dehnen, Astron. J. {\bf 119}, 800 (2000).

34. L. S.Marochnik and A. A. Suchkov, Galaktika (The Galaxy)
(Nauka, Moscow, 1984) [in Russian].

35. T. V. Borkova and V. A. Marsakov, Pis'ma Astron. Zh. {\bf 30},
174 (2004) [Astron. Lett. {\bf 30}, 148 (2004)].

\medskip
{\it Translated by A. Dambis}

\onecolumn
\newpage
\begin{figure}[p]
\begin{itemize}

 {\protect\baselineskip=0.8ex\protect
 {\parshape=1 3.0truecm 12.0truecm
 \item[\bf Table~1.]
 {\protect Parameters of the solar motion.
}

}

}

\end{itemize}
{\begin{center}\parbox{12truecm}{%
\offinterlineskip\halign {#\vrule
 &\hfil#\hfil\vrule
 &\hfil#\hfil\vrule
 &\hfil#\hfil\vrule
   &\hfil#\hfil\vrule
   &\hfil#\hfil\vrule
   &\hfil#\hfil
 &\vrule\hfil#\hfil
 &\vrule\hfil#\hfil\vrule
 &\hfil#\hfil\vrule
 &\hfil#\hfil
 &\vrule#\cr\noalign{\hrule}&&&&&&&&&&&height4pt\cr
 &~ Sample
 &~ $N_\star$
 &~ ${\overline r},$ pc
 &~ ${\overline\tau},$~Gyr
 & $X_\odot$, km/s
 & $Y_\odot$, km/s
 & $Z_\odot$, km/s
 & $V_\odot$, km/s
 & $L_\odot$
 & $B_\odot$
 &\cr
 &&&&&&&&&&&height4pt\cr\noalign{\hrule}&&&&&&&&&&&height4pt\cr
 &~ t1 &~ 1205 ~&~ 107 ~&~ 1.5$\pm$0.3 ~&~11.6$\pm$0.5~&~14.6$\pm$0.5~&~7.0$\pm$0.5~&~19.9$\pm$0.5~&~$52^o\pm 2^o$~&~$21^o\pm 2^o$~&\cr&&&&&&&&&&&height4pt\cr
 &~ t2 &~ 1281 ~&~ 96 ~&~ 2.5$\pm$0.5 ~&~~9.4$\pm$0.6~&~15.2$\pm$0.6~&~7.4$\pm$0.6~&~19.3$\pm$0.6~&~58$\pm$2~&~22$\pm$2~&\cr&&&&&&&&&&&height4pt\cr
 &~ t3 &~ 1304 ~&~ 80 ~&~ 4.3$\pm$1.3 ~&~~7.9$\pm$0.7~&~15.5$\pm$0.7~&~7.8$\pm$0.7~&~19.1$\pm$0.7~&~63$\pm$2~&~24$\pm$2~&\cr&&&&&&&&&&&height4pt\cr
 &~ t4 &~ 1214 ~&~ 69 ~&~ 8.9$\pm$2.5 ~&~~8.6$\pm$0.9~&~25.2$\pm$0.9~&~6.8$\pm$0.9~&~27.5$\pm$0.9~&~71$\pm$2~&~14$\pm$2~&\cr&&&&&&&&&&&height4pt\cr
 \noalign{\hrule}&&&&&&&&&&&height4pt\cr%
 &~ t1$^{'}$ &~ 1116 ~&~ 107 ~&~ 1.5$\pm$0.3 ~&~ 9.7$\pm$0.5~&~13.1$\pm$0.5~&~7.0$\pm$0.5~&~17.8$\pm$0.5~&~53$\pm$2~&~23$\pm$2~&\cr&&&&&&&&&&&height4pt\cr
 &~ t2$^{'}$ &~ 1130 ~&~ 95 ~&~ 2.5$\pm$0.5 ~&~~7.2$\pm$0.5~&~13.8$\pm$0.6~&~6.9$\pm$0.5~&~17.0$\pm$0.5~&~62$\pm$2~&~24$\pm$2~&\cr&&&&&&&&&&&height4pt\cr
 &~ t3$^{'}$ &~~ 968 ~&~ 78 ~&~ 4.2$\pm$1.3 ~&~~5.1$\pm$0.6~&~13.7$\pm$0.6~&~6.7$\pm$0.6~&~16.0$\pm$0.6~&~70$\pm$2~&~25$\pm$2~&\cr&&&&&&&&&&&height4pt\cr
 &~ t4$^{'}$ &~~ 731 ~&~ 67 ~&~ 8.6$\pm$2.5 ~&~~3.0$\pm$0.8~&~16.0$\pm$0.8~&~5.7$\pm$0.8~&~17.2$\pm$0.8~&~79$\pm$3~&~19$\pm$3~&\cr&&&&&&&&&&&height4pt\cr
 \noalign{\hrule}&&&&&&&&&&&height4pt\cr%
 &~ t1$^{''}$ &~~ 71 ~&~ 112 ~&~ 1.6$\pm$0.3 ~&~ 45.9$\pm$2.7~&~34.7$\pm$2.7~&~~8.7$\pm$2.7~&~58.2$\pm$2.7~&~37$\pm$3~&~~9$\pm$3~&\cr&&&&&&&&&&&height4pt\cr
 &~ t2$^{''}$ &~ 102 ~&~ 103 ~&~ 2.5$\pm$0.5 ~&~~36.7$\pm$2.8~&~33.9$\pm$2.8~&~12.2$\pm$2.8~&~51.4$\pm$2.8~&~43$\pm$3~&~14$\pm$3~&\cr&&&&&&&&&&&height4pt\cr
 &~ t3$^{''}$ &~ 186 ~&~ 81 ~&~ 4.5$\pm$1.3 ~&~~18.9$\pm$2.6~&~33.1$\pm$2.6~&~12.3$\pm$2.6~&~40.0$\pm$2.6~&~60$\pm$4~&~18$\pm$4~&\cr&&&&&&&&&&&height4pt\cr
 &~ t4$^{''}$ &~ 294 ~&~ 68 ~&~ 9.3$\pm$2.5 ~&~~16.6$\pm$2.2~&~39.8$\pm$2.2~&~~9.6$\pm$2.2~&~44.1$\pm$2.2~&~67$\pm$3~&~13$\pm$3~&\cr&&&&&&&&&&&height4pt\cr
&\cr
 \noalign{\hrule}
}}
\end{center}
 {\footnotesize
\centerline{$N_\star$ is the number of stars and
 $r$  the heliocentric distance of the star.}

}}
 \vskip 7mm
\end{figure}

\begin{figure}[p]
\begin{itemize}

 {\protect\baselineskip=0.8ex\protect
 {\parshape=1 3.0truecm 12.0truecm
 \item[\bf Table~2.]
 {\protect Parameters of the solar motion

}

}

}
\end{itemize}
{\begin{center}\parbox{12truecm}{%
\offinterlineskip\halign {#\vrule
 &\hfil#\hfil\vrule
 &\hfil#\hfil\vrule
 &\hfil#\hfil
 &\vrule\hfil#\hfil
 &\vrule\hfil#\hfil\vrule
 &\hfil#\hfil\vrule
 &\hfil#\hfil
 &\vrule#\cr\noalign{\hrule}&&&&&&&&height4pt\cr
 &~ 
 & $\sigma_1$
 & $\sigma_2$
 & $\sigma_3$
 & $l_1, ~b_1$
 & $l_2, ~b_2$
 & $l_3, ~b_3$ &\cr &&&&&&&&height4pt\cr\noalign{\hrule}&&&&&&&&height4pt\cr
 &~ t1 &~23.6$\pm$0.6~&~13.7$\pm$0.6~&~10.8$\pm$0.4~&~$15^o\pm 2^o,~~1^o\pm 0^o$~&~$105^o\pm 2^o,~-1^o\pm 1^o$~&~$127^o\pm 2^o,~89^o\pm 4^o$~~&\cr&&&&&&&&height4pt\cr
 &~ t2 &~25.6$\pm$0.7~&~16.0$\pm$0.5~&~13.0$\pm$0.3~&~12$\pm$1,~~0$\pm$0~&~102$\pm$2,~~6$\pm$1~&~281$\pm$2,~85$\pm$4~~&\cr&&&&&&&&height4pt\cr
 &~ t3 &~29.1$\pm$0.8~&~19.2$\pm$0.6~&~15.7$\pm$0.4~&~~7$\pm$1,~~0$\pm$0~&~~97$\pm$2,~~9$\pm$2~&~275$\pm$2,~81$\pm$6~~&\cr&&&&&&&&height4pt\cr
 &~ t4 &~35.4$\pm$1.0~&~23.7$\pm$0.7~&~20.9$\pm$0.6~&~~7$\pm$8,~-1$\pm$1~&~~97$\pm$3,~~8$\pm$2~&~284$\pm$3,~82$\pm$8~~&\cr&&&&&&&&height4pt\cr
 \noalign{\hrule}
}  }%
\end{center}
  }
   \vskip 7mm
\end{figure}

\begin{figure}[p]

\begin{itemize}

 {\protect\baselineskip=0.8ex\protect
 {\parshape=1 3.0truecm 12.0truecm

 \item[\bf Table 3.]
 {\protect
Normalized amplitudes of the main peaks in the UV velocity
distribution

}

}

}

\end{itemize}

{\begin{center}\parbox{12truecm}{%
\offinterlineskip\halign
 {
 #\vrule&\hfil#\hfil&\hfil#\hfil\vrule&\hfil#\hfil\vrule&
 \hfil#\hfil\vrule&\hfil#\hfil\vrule&\hfil#\hfil&\hfil#\hfil&\vrule#\cr
 \noalign{\hrule}&&&&&&&&height4pt\cr
 &~Sample ~&&~ Hyades ~&~ Pleiades ~&~ Sirius ~&~ Coma ~&&\cr
 &&&&&&&&height4pt\cr\noalign{\hrule}&&&&&&&&height4pt\cr
 &~ t1 ~&&~ 48.20 ~&~  8.62 ~&~ 13.88 ~&~  5.84 ~&&\cr&&&&&&&&height4pt\cr
 &~ t2 ~&&~ 17.56 ~&~ 13.42 ~&~ 20.18 ~&~ 11.54 ~&&\cr&&&&&&&&height4pt\cr
 &~ t3 ~&&~ 30.30 ~&~ 11.57 ~&~ 11.59 ~&~  4.31 ~&&\cr&&&&&&&&height4pt\cr
 &~ t4 ~&&~ 16.19 ~&~  2.69 ~&~  5.27 ~&~  3.96 ~&&\cr&&&&&&&&height4pt\cr
 \noalign{\hrule}
} }%
   \vskip 7mm
\end{center}
 }
\end{figure}



\newpage
\begin{figure}[t]
{\begin{center}
 \includegraphics[width= 100mm]{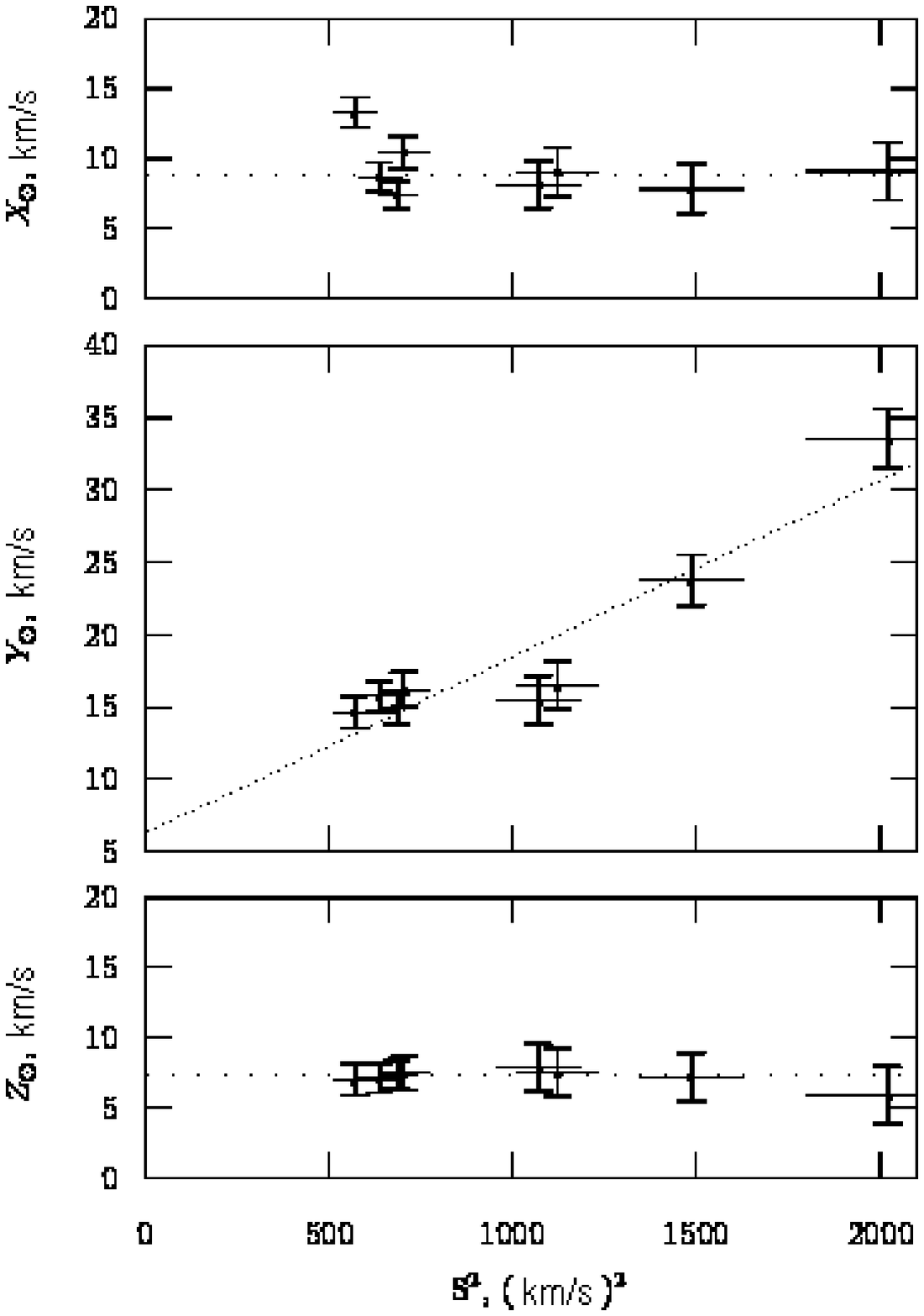}
 \end{center}}
\begin{center} {\bf Fig.~1.} Solar velocity components as a function of the
square of the residual stellar-velocity dispersion.
\end{center}
\end{figure}

\newpage
\begin{figure}[t]
{\begin{center}
  \includegraphics[width= 100mm]{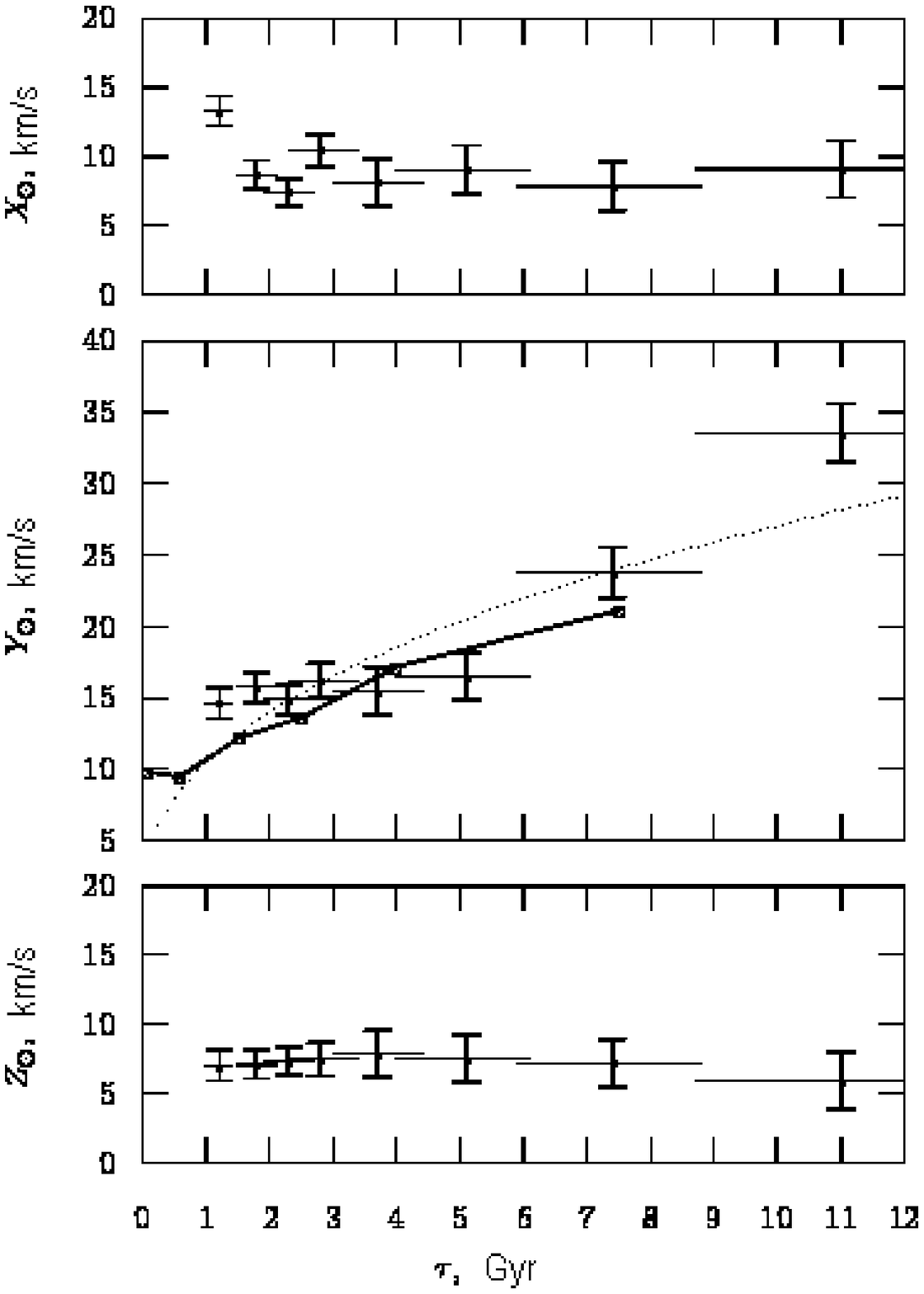}
\end{center}}
\begin{center}{\bf Fig.~2.} Solar velocity components as a function of
stellar age.
\end{center}
 \end{figure}

\newpage
\begin{figure}[t]
{\begin{center}
   \includegraphics[width= 160mm]{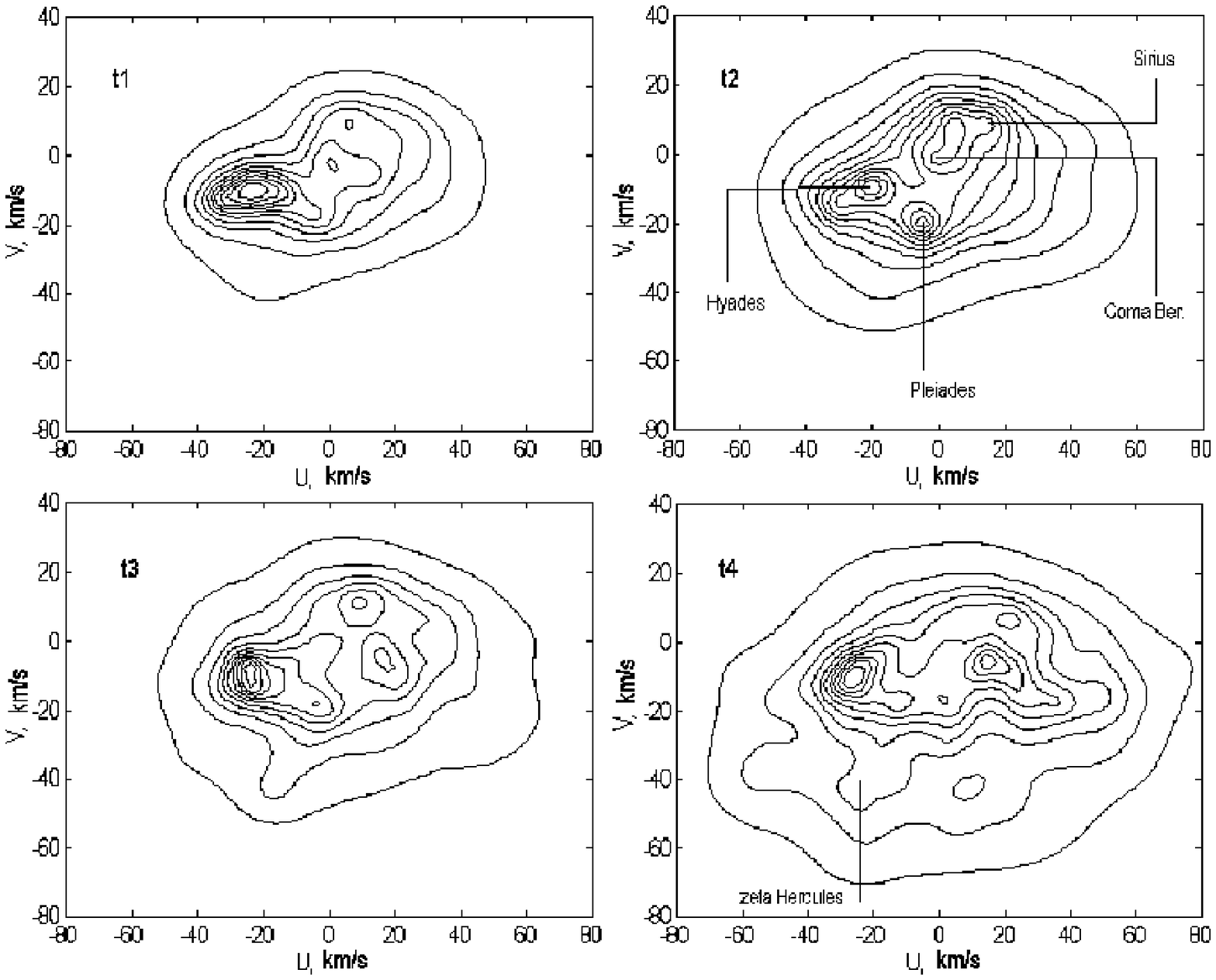}
\end{center}}
\begin{center}
{\bf Fig.~3.} Distributions of $UV$ velocity components of the
stars in four stellar age groups.
\end{center}
 \end{figure}

\newpage
\begin{figure}[t]
{\begin{center}
   \includegraphics[width= 160mm]{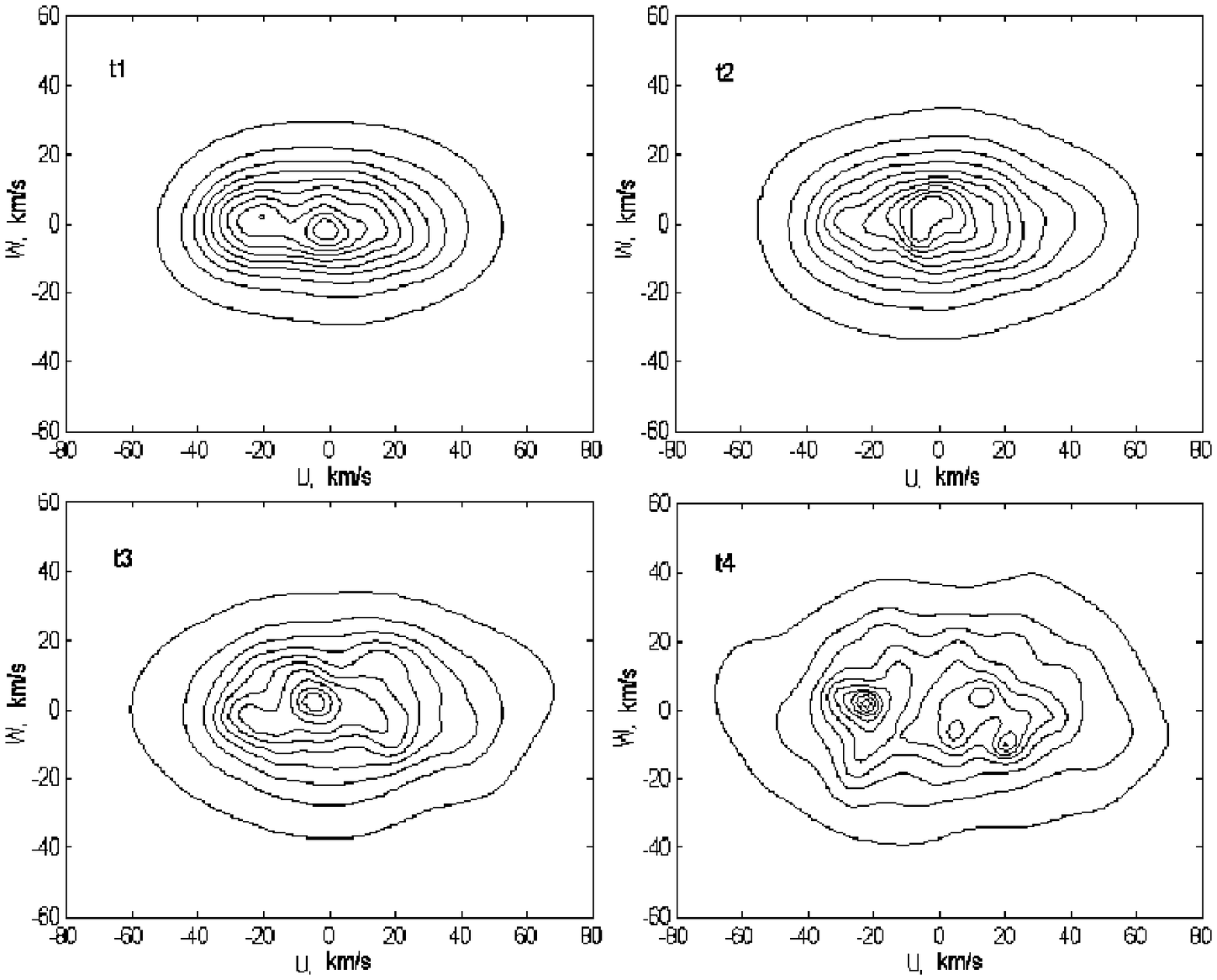}
\end{center}}
\begin{center}
{\bf Fig.~4.} Same as Fig. 3 for the $UW$ velocity components of
the stars.
\end{center}
 \end{figure}

\newpage
\begin{figure}[t]
{\begin{center}
   \includegraphics[width= 160mm]{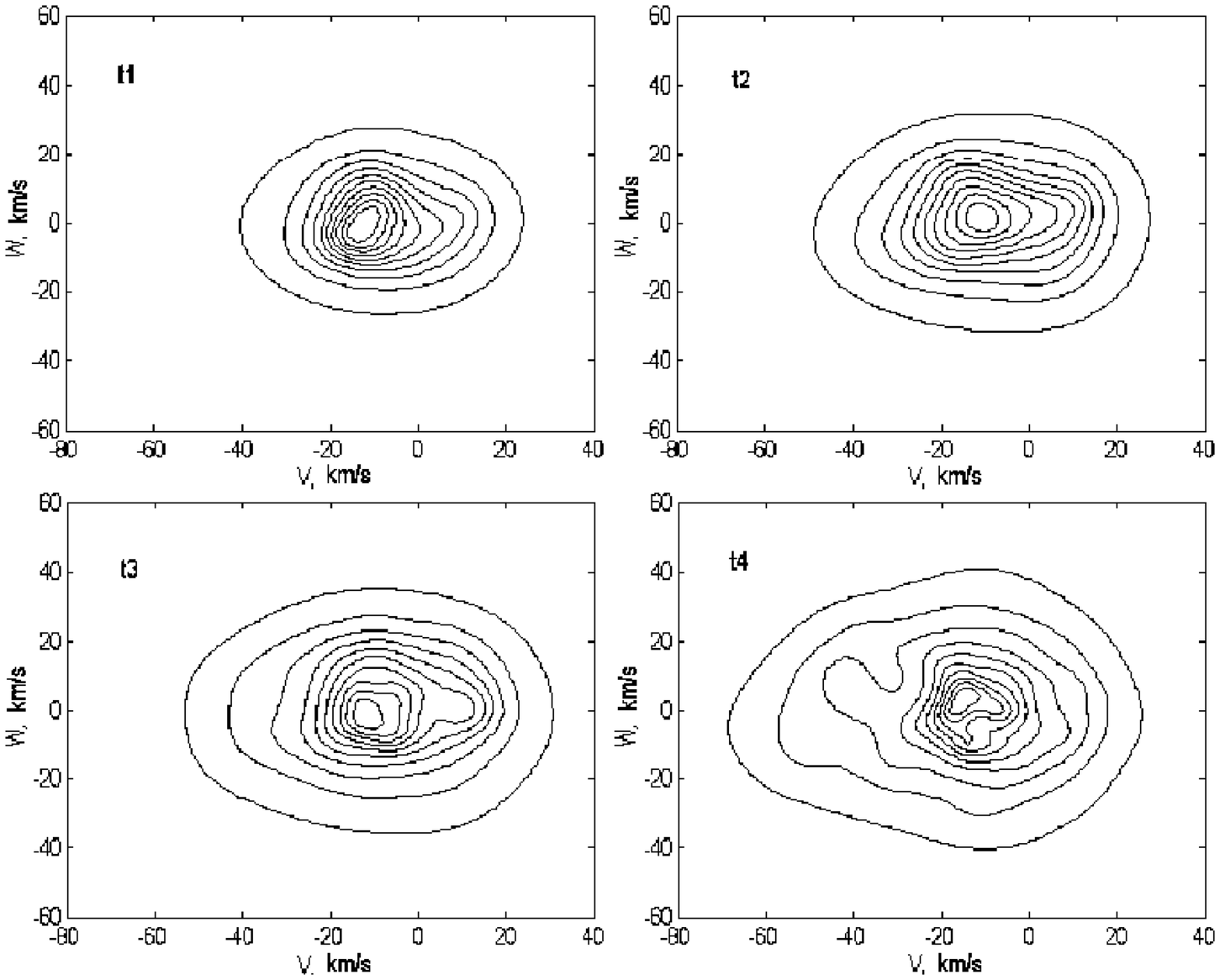}
\end{center}}
\begin{center}
{\bf Fig.~5.} Same as Fig. 3 for the $VW$ velocity components of
the stars.
\end{center}
 \end{figure}

\newpage
\begin{figure}[t]
{\begin{center}
             \includegraphics[width= 100mm]{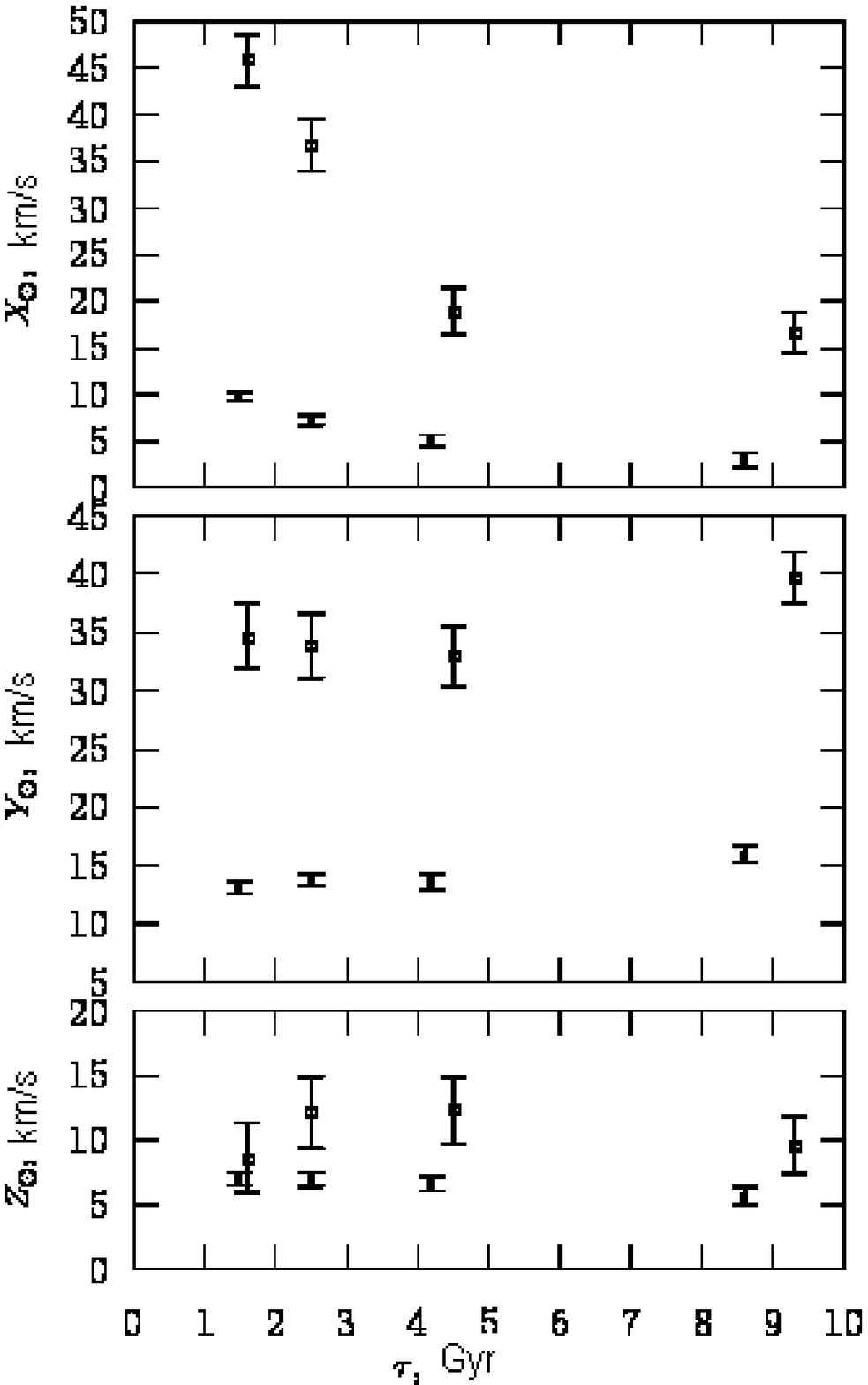}
\end{center}}
\begin{center}
{\bf Fig.~6.} Components of the solar velocity as a function of
the age of stars belonging to the thin disk (filled squares) and
thick disk (open circles).
\end{center}
 \end{figure}

\end{document}